\documentclass{article}

\usepackage{amsmath}
\usepackage{amsfonts}
\usepackage{amssymb}
\usepackage{amsthm}
\usepackage{color}
\usepackage{graphicx}
\usepackage{algorithmic}
\usepackage{algorithm}
\usepackage{authblk}
\usepackage[margin=0.75in]{geometry}
\usepackage{cite}

\newcommand{\LG}[1]{\textcolor{black}{#1}}

\renewcommand{\vec}[1]{{\mathbf{#1}}}

\newcommand{\rev}[1]{#1}
\usepackage[margin=10pt,font=small,labelfont=bf]{caption}

\title{Bayesian Model Averaging for Ensemble-Based Estimates of Solvation Free Energies}

\author[1]{Luke J.~Gosink}
\author[1]{Christopher C.~Overall}
\author[1]{Sarah M.~Reehl}
\author[2]{Paul D.~Whitney}
\author[3]{David L.~Mobley}
\author[4]{Nathan A.~Baker}
\affil[1]{Computational and Statistical Analytics Division, Pacific Northwest National Laboratory, Richland, WA 99352, USA}
\affil[2]{Advanced Computing, Mathematics, and Data Division, Pacific Northwest National Laboratory, Richland, WA 99352, USA}
\affil[3]{Departments of Pharmaceutical Sciences and Chemistry, University of California, Irvine, Irvine, CA 92697}
\affil{Advanced Computing, Mathematics, and Data Division, Pacific Northwest National Laboratory, Richland, WA 99352, USA; Division of Applied Mathematics, Brown University, Providence, RI 02912, USA}

\begin{document}

\maketitle

\begin{abstract}
	This paper applies the Bayesian Model Averaging (BMA) statistical ensemble technique to estimate small molecule solvation free energies.  
	There is a wide range \rev{of} methods \rev{available} for predicting solvation free energies, ranging from empirical statistical models to {\it ab initio} quantum mechanical approaches.
	Each of these methods is based on a set of conceptual assumptions that can affect predictive accuracy and transferability.
	Using an iterative statistical process, we have selected and combined solvation energy estimates using an ensemble of 17 diverse methods from the \rev{fourth Statistical Assessment of Modeling of Proteins and Ligands (SAMPL)} blind prediction study to form a single, aggregated solvation energy estimate. 
	Methods that possess minimal or redundant information are pruned from the ensemble and the evaluation process repeats until aggregate predictive performance can no longer be improved.
	We show that this process results in a final aggregate estimate that outperforms all individual methods by reducing estimate errors by as much as 91\% to $1.2$ kcal mol$^{-1}$ accuracy.
	This work provides a new approach for accurate solvation free energy prediction and lays the foundation for future work on aggregate models that can balance computational cost with \rev{prediction} accuracy.
\end{abstract}

\section{Introduction} \label{Introduction}
Accurate calculation of solvent-solute interactions is an important component of robust molecular simulation including protein structure prediction \cite{Levy:03,Robinson:99, Rakhmanov:07}, conformational ensemble calculations \cite{Jorgensen:2004, Cui:2002, Ashbaugh:99,Ashbaugh:2002}, and binding free energy calculations \cite{Yang:2009, Whalen:2013, Mobley:2009}.
Solvation \rev{free} energy methods for atomically detailed molecular models have a long history of development \cite{Eisenberg:1986, Kang:1987:1, Kang:1987:2, Kang:1987:3,Kang:1987:4, Tan:2006, Gallicchio:2002,Baker:2013,Baker:2015} and have recently benefited from the curation of experimental small molecule solvation data for blind prediction challenges such as the Statistical Assessment of Modeling of Proteins and Ligands (SAMPL) challenge studies \cite{Nicholls:2008, Mobley:2009b, Klimovich:2010, Mobley:2009, Geballe:2012, Geballe:2010, Mobley:2014}.
By presenting a larger range of solvation free energies and molecular weights than seen in common public \rev{datasets}, these challenges are helping to advance the development of solvation methods, \rev{making their estimates more accurate and robust} for a variety of molecular targets~\cite{Ellingson:2014,Muddana:2014,Fu:2014}.  

\rev{Top-performing methods typically come from a wide range of modeling strategies \cite{Nicholls:2008, Mobley:2009b, Mobley:2009, Mobley:2014}}.
Across different challenges, top performers have included explicit solvation methods \cite{Klimovich:2010, Levy:1998, Mobley:2009c}, implicit solvation methods \cite{Mennucci:2007,Jorgensen:2004}, and hybrid methods that combine mixed quantum mechanics (QM) with molecular mechanical (MM) approaches \cite{Konig:2014, Kamerlin:2009}.
In this context, there is a significant degree of uncertainty associated with how to best select, specify, and evaluate the set of parameters and mathematical systems needed to accurately estimate solvation free energies: e.g., uncertainties due to hydrophobicity \cite{Ashbaugh:1999}, surface effects \cite{Chorny:05}, and solvent asymmetries \cite{Mobley:08}.  This type of \emph{method selection uncertainty} affects a wide range of scientific and mathematical disciplines and is arguably the greatest source of error and risk associated with estimation tasks \cite{Rojas:2010, Apostolakis:1990, Devooght:1998, Neuman:2003}.
One of the most powerful ways to address this uncertainty is by combining an ensemble of varied methods (e.g., through a weighted average) to form a single aggregated estimate \cite{Bates:1969, Opitz:1999, Rokach:2010, Hoeting:1999}.
The motivation behind ensemble approaches is based on two principles: (1) most methods in the ensemble possess some unique, useful information; and (2) no single method is sufficient to account for all uncertainties.
When modeled correctly, the information and strengths of individual methods can be combined, and their corresponding weaknesses and biases can be overcome by the strength of the \rev{group}~\cite{Seni:2010, Hoeting:1999,Raftery:1998,Raftery:1995}. 
Ensemble-based estimates are therefore expected to be more reliable and accurate than individual methods, an expectation that has been upheld in numerous examples \cite{Gosink:2014, Zhang:2003, Bates:1969, Morales-Casique:2010, Opitz:1999, Rokach:2010, Hoeting:1999, Seni:2010, Raftery:2005, Vlachopoulo:2013, Raftery:1998, Raftery:1995}.

This paper demonstrates the ability of an ensemble approach called Bayesian Model Averaging (BMA) \cite{Hoeting:1999} to estimate solvation free energies for 45 small molecules by combining predictions from 17 diverse methods provided by the SAMPL4 challenge dataset \cite{Mobley:2014}.
Though BMA has been applied successfully for prediction tasks in many other domains \cite{Ye:2004, Vlachopoulo:2013, Raftery:2005, Morales-Casique:2010, Gosink:2014} this is the first application of the BMA approach \rev{to} solvation energy prediction.

\rev{Throughout this paper, we use the term ``method'' to refer to a specific method (and associated set of parameters, settings, etc.) used to estimate solvation free energies, and the term `model' to refer to some approach for estimating solvation free energies that could be a weighted combination of estimates from one or more methods.
We also define the term ``optimality'', or ``optimal performance'', as the forecasting method that provides the best \emph{overall} performance for targets randomly selected through cross-validation experiments.}

\section{Methods} \label{MethodMain}
\rev{This paper presents a framework for accurately  estimating  small molecule hydration free energies through ensembles of solvation free energy methods.
The paper considers two types of ensembles.
In the simplest case, we examine unique combinations of methods for estimating solvation free energy.
These ensembles of methods are aggregated via a single statistical model to construct an aggregated estimate. 
In the next level of ensembles, we examine multiple such statistical models; i.e. an ensemble of models. 
The framework's objective is to identify the best statistical models, that are in turn based on the best combination of methods.
Through Bayesian Model Averaging, we construct and aggregate an ensemble of ensembles (i.e., multiple statistical models that uniquely combine multiple methods) to create a final aggregated model.
This approach and ensemble have been shown to provide accurate and reliable estimates for numerous problem types~\cite{Gosink:2014, Zhang:2003, Bates:1969, Morales-Casique:2010, Opitz:1999, Rokach:2010, Hoeting:1999, Seni:2010, Raftery:2005, Vlachopoulo:2013, Seni:2010, Hoeting:1999, Raftery:1998, Raftery:1995}.}

\rev{The framework begins by evaluating all solvation methods collectively as an ensemble and uses the ensemble to make an aggregated estimate of free energy. 
The framework then performs an iterative process of ensemble pruning, analyzing, and estimating.
The pruning process uses statistical models that collectively assess the information content of each method.
During each iteration, this pruning process removes the most redundant or least accurate methods from the ensemble.
After pruning, performance for the new ensemble is assessed through a set of cross-validation studies.
The ensemble that provides the best overall performance is defined as optimal and is the best choice for future estimation tasks (e.g., for future SAMPL challenges).}

\rev{When an aggregate estimate proves to be more accurate than individual methods, understanding how individual methods were combined can help improve method development.
Towards this objective, this framework provides quantitative information to characterize the optimal ensemble by answering the following types of questions.
Did the ensemble relied solely on implicit or explicit methods, or if it used variety of different methods?
Which methods contributed the most useful information in the aggregate estimate?
Alternatively, if the ensemble is largely composed of one type of method, how do variations in parameters influence the results?}

\subsection{Model specification with Bayesian Model Averaging (BMA)} \label{Method} 

\rev{The simplest BMA approach assembles a set methods into a linear system}~\cite{Hoeting:1999,Raftery:1998,Raftery:1995}.  
Let $y_i$ for $i = 1, \ldots, N$ be a series of hydration free energy observations for a collection of $N$ molecules, and let $x_{i j}$ denote the \rev{free energy estimate for the $i^{th}$ molecule} obtained from \rev{$1 \leq j \leq P$ prediction methods. 
The $x_{ij}$ values form a numerical ensemble estimate matrix} that, along with $y_i$, defines a linear regression model 
\begin{equation}
	\label{Method:E1}
	y_i = \sum_{j=1}^P x_{ij} \;\beta_j + \epsilon_i
\end{equation}
\rev{where the parameter vector $\beta_j$ defines the unknown relationship between the ensemble's $P$ constituents and $\epsilon_i$ is a disturbance term that captures all factors that influence the dependent variable $y_i$ \emph{other} than the regressors $x_{ij}$.
Such factors include latent variables impacting the response $y_i$, nonlinearity effects, and measurement errors associated with quantifying $y_i$, as well as stochastic, experimental effects that impact the ability to reproduce $y_i$.  
Explicitly solving for the disturbance term is not feasible: $\epsilon_i$ is a theoretical, non-observable random term that must be estimated through an analysis of model residuals.  
Fortunately, if the distribution of $\epsilon_i: i=1,\ldots,N$ is normal (i.e., $\epsilon \sim \mathcal{N}(0,\sigma)$) the total disturbance term's contribution to the model's estimate goes to zero as the expectation of random variables with such distributions is $E [\epsilon | X]\ = 0$.}

\rev{Under this assumption of normality, Equation~\ref{Method:E1} estimates the values $\beta_j$ that will fit both the known hydration free energy data in $y_i$ and facilitate the ability to make inferences on the hydration free energy of unknown molecules.  
Many different regression techniques can estimate $\beta_j$ \cite{Hosmer:1989,Reiss:2012,Mallows:1973,Candes:2007}; however, these techniques commonly generate estimates that vary in their ability to model and infer \cite{Genell:2010,Hoeting:1999,Davidson:2006,Raftery:1995,Raftery:1998}.}

BMA addresses the challenge of statistical model uncertainty by first evaluating all $2^{P}-1$ possible statistical models that can be formed from the $P$ estimation methods and combining each model's $\beta_j$ through a weighted average into an aggregated parameter vector, $\beta_j^{\text{BMA}}$ (Equation~\ref{Method:E2}).
There are $k = 1, \ldots, 2^P-1$ distinct combinations of the $P$ estimation methods, each with a corresponding statistical model, $M^{(k)}$, and parameter vector, $\beta^{(k)}_j$. 
BMA combines each $\beta_{j}^{(k)}$ through an average that weights each $\beta^{(k)}_j$ by the probability that its statistical model, $M^{(k)}$, is the ``true'' model:
\begin{equation} \label{Method:E2}
	\LG{\beta_j^{\text{BMA}} = E[\,\beta_j \,|\, \+y\,]  = \sum_{k=1}^{2^P-1} E[\beta_{j}^{(k)}\,|\, \+y,M^{(k)}\,] \, {\mathrm{Pr}}(M^{(k)} | \+y)}
\end{equation}
where $E[\beta_{j}^{(k)}\,|\, \+y,M^{(k)}\,]$ is the expected value of the posterior distribution of $\beta^{(k)}_j$.
This distribution is weighted by the posterior probability ${\mathrm{Pr}}(M^{(k)} | \+y)$ that $M^{(k)}$ is the \emph{true} statistical model given the data $\textbf{y}$.
The expected posterior distribution of $\beta^{(k)}_j$ is approximated through the linear least squares solution of the given model $M^{(k)}$ and solvation energy response variable $\vec{y}$.
The posterior probability term is estimated from information criteria \cite{Raftery:1995}
\begin{equation} \label{Method:E3} 
	{\mathrm{Pr}}(M^{(k)} | \+y) \propto \frac{e^{-\frac{1}{2}B^{(k)}}}{\sum^{2^P-1}_{l=1} e^{-\frac{1}{2}B^{(l)}}} 
\end{equation}
where $B^{(k)}$ is the Bayesian Information Criteria for model $M^{(k)}$ estimated as
\begin{equation} \label{Method:E4} 
	B^{(k)} \approx N \log{(1-R^{2(k)})} + p^{(k)} \log{N}.
\end{equation}

\rev{The Bayesian Information Criteria (BIC) is a method for ranking models based on goodness of fit, $R^2$.
The $R^2$ is a statistical measure of how close the data are to the fitted regression line, essentially the percentage of the response variable variation that is explained by a linear mode.
The BIC strongly penalizes models for having too many parameters, striking a parsimonious balance between sufficient complexity for a model to be useful and over-fitting the data. 
In Equation~\ref{Method:E4},  $R^{2(k)}$ is this adjusted $R^2$ for model $M^{(k)}$ that indicates the model's goodness of fit for the observations, $p^{(k)}$ is the number of methods used by the model (not including the intercept), and $N$ is the number of solvation free energy values to be predicted (i.e., the number of molecules).}

The parameter vector $\beta_j^{\text{BMA}}$ obtained from Equation~\ref{Method:E2} helps to address model uncertainty by accounting for all systems of linear equations that can model the relationship between the measured solvation free energy values $y_i$ and values $x_{ij}$ predicted by each solvation method $j$.
Over-fitting by more complicated models is addressed through the Bayes information criterion, which penalizes models with more variables \rev{(i.e., statistical models that use more methods in their regression)}.
Perhaps more importantly, $\beta_j^{\text{BMA}}$ can be used to estimate new solvation free energy values for unmeasured molecules by combining new $x_{i j}$ estimates.

\subsection{\rev{Pruning models with Occam's window}} \label{Method:StatEnsemble}

The inclusion of all $2^P-1$ ensemble models in Equation~\ref{Method:E2} is not necessarily beneficial for predictive performance.
\rev{A ``misspecified'' ensemble model may perform poorly if it omits a method (and resultantly information) that is crucial to estimating $y_i$.
Alternatively, a model may perform poorly if it includes methods that provide inaccurate information about $y_i$.
Despite the fact that they are down-weighted via low posterior probabilities, the cumulative effect of these misspecified models, can erode the ensemble's overall performance \cite{Qian:2015, Martinez-Munoz:2009, Raftery:1998, Onorante:2014, Madigan:1994, Hoeting:1999, Morales-Casique:2010}.
Madigan and Raftery~\cite{Madigan:1994} developed the Occam's Window ensemble-pruning approach that eliminates under-performing models based on the BIC discussed above.}
Occam's Window defines a set of models
\begin{equation} \label{Method:E5}
	\textbf{A} := \{M^{(k)} \in \textbf{M} | BIC^{(k)} - BIC^{(min)} < 6\},
\end{equation}
where $BIC^{(min)}$ denotes the BIC of the model $M^{(k)}$ with the lowest $BIC$ \rev{(highest information content)}.
The value 6 is based on Jeffreys' \cite{Jefferys:1961} and Raftery's \cite{Raftery:1995} assessment of Bayes factors for model comparison and ensures all $m_k \in \mathbf{A}$ meet a minimum statistical information criteria for the aggregation process.
Constraining Equation~\ref{Method:E5} to the set $\+A$ accelerates the evaluation of Equation~\ref{Method:E5} and improves BMA's predictive capability \cite{Raftery:1998, Madigan:1994}:
\begin{equation} \label{Method:E6}
	\beta^{BMA}_j =  E[\,\beta_j \,|\, \+y\,] = \sum_{M^{(k)}\in \mathbf{A}} E[\,{\beta_j}^{(k)} \,|\, \+y,M^{(k)}\,] \,Pr(M^{(k)} \,|\, \textbf{y})
\end{equation}

BMA estimates method $j$'s utility for explaining a set of observations, \textbf{y}, by assessing the probability that the method's coefficient term, $\beta^{BMA}_j$ will receive a non-zero value.
The estimate of this probability is based on the conditioned, cumulative sum of all model posteriors
\begin{equation} \label{Method:E7} 
	\mathrm{Pr}(\beta_j^{\text{BMA}}\neq 0) =   \sum_{M^{(k)}\in\mathbf{A}} {\mathrm{Pr}}(M^{(k)} | \textbf{y}) \:{\mathrm{I(j)}}
\end{equation}
where
\begin{equation} \label{eq:bma-prob-neq0-ID}
	\mathrm{I(j)}: =
	\begin{cases} 
		1,&\text{if model $M^{(k)}$ specifies j as a regressor} \\
		0,&\text{ otherwise}.
	\end{cases}
\end{equation}
Equation~\ref{Method:E7} \rev{sorts and prioritizes} the methods based on their utility; i.e., the probability that the coefficient term weighting a method's estimate will \emph{not} be 0.
The combination of Equations~\ref{Method:E6} and ~\ref{Method:E7} therefore provide a statistical framework to support an iterative, statistical process for designing an ensemble that will provide the best combination of models and methods for estimating solvation free energy.

\subsection{\rev{Statistical design of method ensembles}}
 \label{Method:StatEnsemble:Design}
\rev{The statistical design process, shown in Algorithms~\ref{Method:Alg1} and~\ref{Method:Alg2}, proceeds in two iterative stages. 
In both algorithms, we denote matrix and vectors with bold script and indicate row and column dimensions in superscript.
For example, $\textbf{x}^{n \times p}$ is a matrix with $n$ rows and $p$ columns, where as $x$ is a scalar, or real value.
Additionally,  $\textbf{0}^{n \times p}$ is a zero matrix and  $\textbf{x}^{1 \times p}$ is a zero vector; essentially, a matrix and vector of the specified dimensions where all values are 0.}

\rev{The first stage, presented in Algorithm~\ref{Method:Alg2}, uses Equation~\ref{Method:E7} to evaluate the statistical significance of each method in the current version of the ensemble.
This information is based on a set of 2-fold cross-validation tests that randomly create training and validation datasets from the small molecule compounds  in the SAMPL4 challenge.
In these tests, each method's statistical information is thus evaluated against a variety of random training and validation sets.
The results of this first stage are then sent to stage two (Algorithm~\ref{Method:Alg1}): the ensemble pruning stage.}

\rev{The second stage evaluates the performance of the current ensemble and then prunes this ensemble by removing that method with the least statistical significance.
The process then repeats using the new pruned ensemble to identify the ideal ensemble of methods.}

\rev{At a more detailed level, the design process assumes an initial set of $n$ estimates made by $p$ methods, $\textbf{x}^{n \times p}$ for observations, $\textbf{y}^{n \times 1}$.
The second and third lines initialize the root mean squared error (RMSE) and the solution for the best ensemble of models and methods.
The algorithm will iterate over the following tasks while it still has more than two methods to evaluate (lines 5 -- 19).}

\begin{table}[t] 
	\begin{minipage}[t]{0.45\linewidth}\centering 
		\begin{algorithm}[H] 
			\footnotesize 
			\caption{\newline \rev{Prune and Design Ensemble}} \label{Method:Alg1} 
			\begin{algorithmic}[1] 
				{\scriptsize 
				\REQUIRE \textbf{vector} $\textbf{y}^{n \times 1}$, \textbf{matrix} $\textbf{x}^{n \times p}$\; 
				\vspace{2.5mm} 
				\STATE $loopCount = 1$\; 
				\STATE $rmse = null$\; 
				\STATE $\textbf{x\_solution} = \textbf{x}$\; 
				\STATE $\boldsymbol{\beta}_{BMA}^{1 \times j} = \textbf{0}^{1 \times j}$ \COMMENT{initialize coefficients to zero}\; 
				\WHILE{$($loopCount $<$ p$)$}\; 
					\STATE $\boldsymbol{\beta}_{BMA}, s, \textbf{v} = \textbf{Analyze Methods}(\textbf{y, x})$\; 
					\IF{$($s $<$ rmse$)$ or $($rmse == null$)$}\; 
						\STATE $\textbf{x\_solution} = \textbf{x}$\; 
						\STATE $rmse = s$\; 
					\ENDIF\; 
					\STATE $m = 1$\; 
					\FOR{$($$methods$ $\in$ \textbf{x}$)$}\; 
						\IF{$($\textbf{v}[$methods$] $<$ \textbf{v}[$m$]$)$}\; 
							\STATE $m = methods$\;
						\ENDIF\;
					\ENDFOR\;
					\STATE $\textbf{x} = $\textbf{x}.delete($m$)
					\COMMENT{prune least significant method}\;
					\STATE $loopCount += 1$\;
				\ENDWHILE\;
				\RETURN $\textbf{x\_solution},\boldsymbol{\beta}_{BMA}$, rmse\;} 
			\end{algorithmic}
		\end{algorithm}
	\end{minipage}
	\hspace{0.5cm}
	\begin{minipage}[t]{0.45\linewidth}
		\centering
		\begin{algorithm}[H] 
			\footnotesize 
			\caption{\newline \rev{Analyze Methods}} \label{Method:Alg2} 
			\begin{algorithmic}[1] 
				{\scriptsize 
				\REQUIRE  \textbf{vector} $\textbf{y}^{n \times 1}$, \textbf{matrix} $\textbf{m}^{n \times j}: j \leq p$\; 
				\vspace{2.5mm} 
				\STATE $cv = 100$\; \COMMENT{100 cross-validation experiments}\; 
				\STATE $\boldsymbol{{\beta}}_{BMA}^{1 \times j} = \textbf{0}^{1 \times j}$ \COMMENT{initialize coefficients to zero}\; 
				\STATE $error = 0$ \COMMENT{mean RMSE of the cross-validation}\; 
				\STATE $\textbf{v}^{1 \times j} = \textbf{0}^{1 \times j}$ \COMMENT{vector storing each method's p$\neq$0}\;  
				\WHILE{$($$index$ $<$ $cv$$)$}\; 
					\STATE $\textbf{m}_{t}^{q \times j}, \textbf{m}_{v}^{r \times j} =\ \textbf{m}^{n \times j}$\;
					\COMMENT{split $\textbf{m}$: train and test}\; 
					\STATE $\boldsymbol{\beta}_{BMA} = \boldsymbol{\beta}_{BMA} + $BMA$(\textbf{m}_{t})$ \COMMENT{Equation~\ref{Method:E6}}\; 
					\STATE $error = error + RMSE(\textbf{Y}, \textbf{m}_{v} \times \boldsymbol{\beta}_{BMA})$\; 
					\STATE $\textbf{v} = \textbf{v} + $P$\neq$0$(\textbf{m}_{t})$ \COMMENT{Equation~\ref{Method:E7}}\; 
					\STATE $index = index + 1$ 
				\ENDWHILE\; 
				\RETURN $\frac{\boldsymbol{\beta}_{BMA}}{cv}$,$\frac{error}{cv}$,$\frac{\textbf{v}}{cv}$\; } 
			\end{algorithmic}
		\end{algorithm} 
	\end{minipage}
\end{table}

First, the performance of the current ensemble is evaluated through \emph{Analyze Methods} (line 6).
\rev{The \emph{Analyze Methods} function in Algorithm \ref{Method:Alg2} takes as input the same observation data $\textbf{y}$ as Algorithm \ref{Method:Alg1} as well as a matrix  $\textbf{m}^{n \times j}$ that contains estimates made by a subset of $j \leq p$ methods.
The matrix $\textbf{m}^{n \times j}$ represents the current set of methods from $\textbf{x}^{n \times p}$ still being analyzed.}
Algorithm \ref{Method:Alg2} performs 100 iterations of a 2-fold cross-validation and begins by partitioning the estimate matrix $\mathbf{m}$ equally into training and validation data (line 6).
Using the training data, the algorithm uses Equation~\ref{Method:E6} to estimate $\beta_{BMA}$ and then calculates the RMSE of this coefficient vector based on the validation data (lines 7 and 8).
Next, the $\mathrm{Pr}(\beta_j^{\text{BMA}}\neq 0)$ of each method are calculated and saved in the vector, \textbf{v} (line 9).
\rev{After 100 iterations, the function returns an estimate for $\beta_j^{\text{BMA}}$, the mean RMSE of the current ensemble, as well as a list of all methods and their corresponding probability values for $\mathrm{Pr}(\beta_j^{\text{BMA}}\neq 0)$.}
Next, Algorithm \ref{Method:Alg1} compares and conditionally updates the current ``best'' performing model, \textbf{x\_solution}, with the model returned by \emph{Analyze Methods} (lines 8 and 9).
The algorithm then identifies the method with the lowest $\mathrm{Pr} (\beta_j^{\text{BMA}}\neq 0)$ and removes this method from the current model (line 17). With this method removed, the process repeats until there are just two methods left.
The output of this process is an ensemble of methods, $\textbf{x}_{ij}$ that are statistically determined to be the best methods for estimating the observations in $\mathbf{y}$ and a statistical model, $\beta_{BMA}$ that specifies how to best combine these methods.  

\subsection{Solvation free energy data and solvation methods} \label{EP:DataModels}
The SAMPL4 challenge consists of 49 submissions representing a total of 19 different research groups \cite{Mobley:2014}.
Each of these methods provides solvation free energy estimates for 45 small molecule compounds.\footnote{\rev{The original SAMPL4 study used 52 samples; thus some samples have labels like SAMPL4\_052.
However, several compounds were removed during the course of the study based on ``problems with experimental values, SMILES strings, or structures for a number of compounds, resulting in removal of some compounds from the challenge.''\cite{Mobley:2014}}}
The challenge is blind in the sense that the solvation free energy values for these molecules were hidden from participants; e.g., free energy values are not found in standard solvation free energy test sets, and their values are not readily available in the literature \cite{Guthrie:2014}. 
In this work, we restrict our analysis to a subset of 17 method submissions based on the fact that many groups made multiple submissions that were strongly correlated.
In these cases, we chose only a single variant to ensure that multicollinearity did not inflate the significance of specific methods during model selection and averaging; such bias can create unstable estimates for $\beta^{BMA}_j$ and that \rev{in turn} can reduce BMA's \rev{estimate} accuracy \cite{Clyde:1999}.
These methods are summarized in Table~\ref{Analysis:Table1:Methods} based on the methods used to calculate the solvation energy: 
\begin{itemize} 
	\item \textbf{Group 1:} Single-conformation implicit solvent methods \cite{Ellingson:2014, Nicholl:2010, Hawkins} based on Poisson-Boltzmann and related methods \cite{Fixman:1979, Honig:1995, Davis:1990};
	\item \textbf{Group 2:} Multi-conformational implicit solvent methods~\cite{Sandberg:2013, Klamt:2009, Hogues:2014, Sulea:2011, Reinisch:2014}; 
	\item \textbf{Group 3:} Methods based on molecular dynamics-based free energy calculations in explicit solvent \cite{Klimovich:2010,Muddana:2014,Mobley:2009c,Mobley:2007} with small molecule force fields \cite{Wang:2004B};  and, 
	\item \textbf{Group 4:} Hybrid solvent methods \cite{Li:2014}.
\end{itemize}
\begin{table}[t]
	\centering
	\caption[Ensemble Constituents]{This table lists the solvation methods used in our ensemble design process.
	Method ID indicates the identification number of the method that is referenced throughout this paper.
	\rev{CPU time is an order-of-magnitude estimate of the method's reported time to make a prediction.}
	Sampling strategies include quantum mechanical (QM), molecular dynamics (MD), and molecular mechanics with Poisson-Boltzmann surface area solvation (MM-PBSA).
	The listed performance is based on 100 iterations of a 2-fold cross-validation study as described in the text.
	This performance is also shown graphically in Figure~\ref{Analysis:Figure1:Methods}.
	The last column is a comparison of each method to the optimal BMA ensemble, BMA (Stage 16).
	This column indicates that the ensemble design approach presented in this work is able to reduce estimation errors by 29\% to 91\% (i.e., 1.2--9.4 kcal mol$^{-1}$) in comparison to the individual methods.
	The final BMA ensemble is indicated by the two italicized methods: imp-2 and alc-3.
	Wilcoxon based $p$-values for BMA's mean RMSE distribution vs.\@ the best method's mean RMSE distribution (imp-2) are shown in Table~\ref{Analysis:Table2:BMA}.}
	\scriptsize
	\rev{
	\begin{tabular}{l|r|c|c|c|c}
		\hline
		\hline
		Reference 	& Method	ID & CPU Time & Methodology  &Ensemble Mean				& Improvement \\
					&  		&			&	& RMSE (kcal mol$^{-1}$)	& with BMA  \\
					& 			&		&		& and Std Err Mean	& (Stage 16) \\
		\hline
		Coleman et al.~\cite{Coleman:2014} & imp-6  & seconds & multi-conformation implicit & 9.40 $\pm$ 0.2 & 91\% \\
		Parsod et al. & alc-5 & hours & alchemical MD & 4.50 $\pm$ 0.071 & 82\% \\ 
		Sharp et al.~\cite{Yang:2006} & imp-4 &minutes& single conformation implicit & 3.80 $\pm$ 0.045 & 78\% \\
		Jiafu et al. & exp-3 &hours& QM/MD & 2.80 $\pm$ 0.045 & 71\% \\
		Purisima et al.~\cite{Hogues:2014} & imp-5&minutes  & single conformation implicit & 2.55 $\pm$ 0.049 & 68\%\\
		Mark et al.~2011 & alc-2 & hours&alchemical MD & 2.40 $\pm$ 0.036 & 66\%\\ 
		Genheden et al.~\cite{Genheden:2014} & exp-1&hours  & MM-PB/SA & 2.00 $\pm$ 0.022 & 59\%\\
		Weyang et al. & exp-4 &hours & MM-PB/SA & 1.84 $\pm$ 0.031 & 55\%\\		
		Biorga et al.~\cite{Beckstein:2012,Beckstein:2014} & alc-4& hours & alchemical MD & 1.65 $\pm$ 0.013 & 50\%\\ 
		Elingson et al.~\cite{Ellingson:2014} & imp-7 &seconds & single conformation implicit & 1.52 $\pm$ 0.039 & 46\%\\
		Fennell et al.~\cite{Li:2014} & hyb-2 & hours & hybrid & 1.52 $\pm$ 0.013 & 46\%\\																
		Jambeck et al.~\cite{Jambeck:2013} & alc-1&hours  & alchemical MD & 1.52 $\pm$ 0.020 & 46\%\\ 
		Park~\cite{Park:2014} & imp-3 &seconds  & single conformation implicit & 1.44 $\pm$ 0.020 & 43\%\\
		Klamt et al.~\cite{Klamt:2010} & imp-1 &hours & single conformation implicit & 1.36 $\pm$ 0.030 & 40\%\\	
		\textit{Gilson et al.~\cite{Muddana:2014}} & \textit{alc-3} & hours & \textit{alchemical MD} & {1.24 $\pm$ 0.017} & {34\%} \\ 
		Geballe et al.~\cite{Ellingson:2014} & imp-8 & seconds & single conformation implicit & 1.17 $\pm$ 0.017 & 30\%\\
		\textit{Sandberg et al.~\cite{Sandberg:2013}} & \textit{imp-2} &minutes  & \textit{multi-conformation implicit} & \textit{1.15 $\pm$ 0.023} & \textit{29\%}\\
		\textbf{BMA (Stage 16)} & \textbf{NA} & seconds & \textbf{ensemble} & \textbf{0.82} $\pm$ \textbf{0.015} & \textbf{0} \\
		\hline
	\end{tabular}}
	\label{Analysis:Table1:Methods}
\end{table}
\begin{figure}[h!]
	\centering
	\includegraphics[keepaspectratio,width=0.9\textwidth]{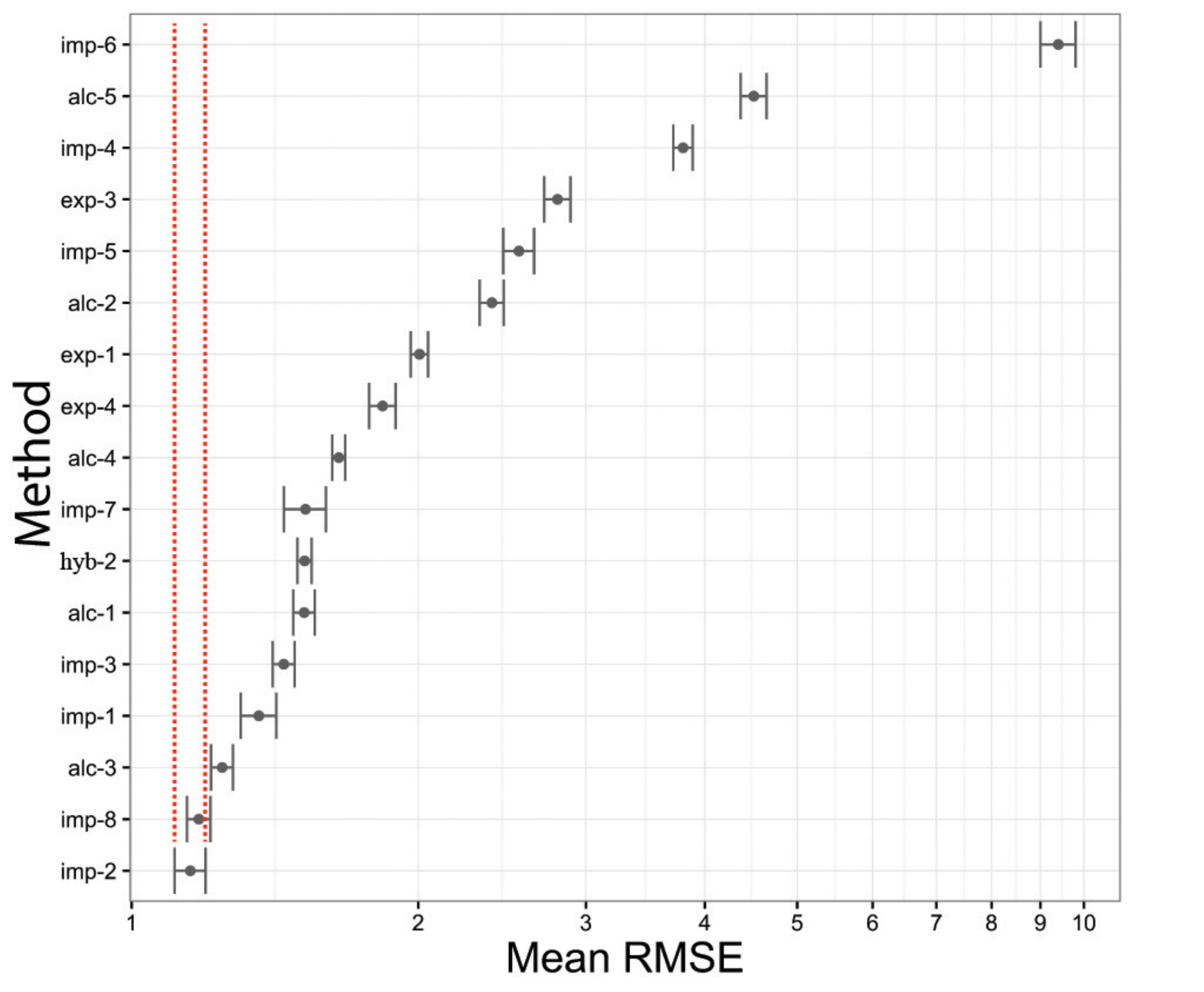}
	\caption{
	This figure uses a squadron of Imperial Tie Fighters \cite{Lucas:77} to depict the mean root mean squared error \rev{(RMSE)} for the 17 initial methods used in our ensemble design process.
	\rev{Additionally, the 95 \% confidence interval (based on standard error) for the expected mean RMSE is shown for each method.
	Thus at 95\% statistical confidence, the true expected performance for each method is depicted according to the 100 iterations of the 2-fold cross-validation experiment we detail in Algorithm 2.}}
	\label{Analysis:Figure1:Methods}
\end{figure}
 
 \subsection{\rev{Training, estimating, and pruning an ensemble with BMA}} \label{EP:training}
From this initial set of 17 methods, we iteratively assessed and pruned methods as described in Algorithm \ref{Method:Alg1}.
The approach began by randomly sampling (without replacement) 26 of the original 52 experimental solvation free energy measurements to form training and validation datasets (Algorithm \ref{Method:Alg2}, line 6).
Collectively, these sampled values formed the observation vector $y_i$ and the estimates from each of the prediction methods in Table~\ref{Analysis:Table1:Methods} for these measurements defined the ensemble estimate matrix, $x_{i j}$.
The observation vector and the ensemble estimate matrix were used to form the linear system in Equation~\ref{Method:E1}.  

Next, we estimated the $\beta_j^{\text{BMA}}$ parameter from Equation~\ref{Method:E2} by assessing all possible ($2^{17} -1$) statistical models $M^{(k)}$.
Each model's information criteria $B^{(k)}$ was used to identify a reduced set of \rev{the} most informative models per Equation~\ref{Method:E5}.
Based on this reduced modeling space, the coefficient terms $\beta_j^{\text{BMA}}$ were estimated through weighted averages of each statistical model's ordinary least squares solution (Equation~\ref{Method:E6}).
We used $\beta_j^{\text{BMA}}$ to estimate the remaining 26 solvation free energy measurements that were \emph{not} used to train the BMA model.
This task was accomplished by combining the estimates of all methods in Table~\ref{Analysis:Table1:Methods} for the validation data with $\beta_j^{\text{BMA}}$ to produce an aggregated estimate. 
A \rev{RMSE} was obtained for the 26 estimates made on validation data.

This process was repeated 100 times (Algorithm \ref{Method:Alg2}) so that performance for any estimation method could be reported as a mean RMSE: i.e., the mean of 100 RMSEs that each represent performance for a 2-fold cross-validation using the 26 member validation \rev{sets} described above.
In addition to the mean RMSE, the information content provided by each method in the ensemble is also returned. 

Algorithm \ref{Method:Alg1} used this information to perform two tasks.
First, if the performance of this aggregated estimate was better than all previously examined aggregates, the statistical model combining these methods was saved as the new optimal model (Algorithm \ref{Method:Alg1}, lines 7-9).
Next, Algorithm \ref{Method:Alg1} pruned \rev{a} method from the ensemble that provided the least amount of information to the aggregated estimate.
The pruning process was repeated until just two methods are left (Algorithm \ref{Method:Alg1}, lines 12-18).
The ensemble of methods whose aggregate forecast has the lowest mean RMSE was saved and returned as the final model for the ensemble.
This final model, and the set of methods that correspond to this model, were the final products of the statistically driven ensemble design process, referred to as the BMA-based optimal ensemble.

\section{Results and Discussion} \label{Results}
We report the results of our \rev{BMA-based ensemble} in three stages.
First, we contrast the performance of the initial 17 SAMPL4 methods to the performance of the BMA-based optimal ensemble.
We then present data on the iterative design process used to create this optimal ensemble.
Second, we examine the \emph{conditional} performance of this optimal ensemble according to the individual molecules used in the SAMPL4 challenge.
In this stage, we show how the BMA-based optimal ensemble provides more reliable estimates in comparison to individual methods, especially against the more challenging set of small molecules from the SAMPL4 dataset.
Finally, we complete the analysis of the optimal ensemble by contrasting its performance to the performance of alternate statistical techniques that can be used to form aggregated estimates. 

\rev{In all analyses, we report the performance based on the mean RMSE associated with the 100 $\times$ 2-fold cross-validation. 
Further, we report the standard error of the mean (SEM) to better characterize the expected performance for these estimates. 
In contrast to standard deviation, which quantifies the spread of the cross-validation distribution, the SEM metric directly maps cross-validation performance into confidence intervals that bound the true expected performance of each RMSE.}
 
We report the statistical significance of all performance data through a Wilcoxon rank sum paired comparison test \cite{Wilcoxon:45}.
This non-parametric approach tests the hypothesis that the mean RMSE distributions of two approaches are equal: $H_0 : \mu_Y = \mu_X$.
Thus when comparing BMA to a given method, a Wilcoxon generated $p$-value greater than 0.05 indicates we fail to reject $H_0$: the distributions are thus equal and we conclude that BMA and the method are equivalent in their performance.
On the other hand, Wilcoxon-generated $p$-values that are less than 0.05 indicate we should reject $H_0$.
In this latter case, we then compare the mean RMSE for BMA and the given method to assess performance.

\rev{To control the family-wise error rate of our tests, we applied a Bonferroni correction to determine $p$-values with a threshold of $\alpha = 0.05$.
This correction addresses a problem that arises in statistical analysis for a large number of concurrent comparisons.
Larger comparison spaces are more likely to yield a statistically significant observation by chance, resulting in a false positive.
Bonferroni corrections adjust the false positive rate to account for this effect.}

Finally, a Shapiro-Wilk test~\cite{Shapiro:65} was used to test the hypothesis, $H_{0}$, that residuals in the SAMPL4 challenge are not significantly different from a normal population.
This test is used to determine if samples from our distribution or residuals are distributed normally.
Thus, under an $\alpha = 0.05$, a p-value statistic of 0.9556 was obtained from this test, indicating that there was insufficient information to reject $H_{0}$. 
Without evidence to reject $H_{0}$, assumptions of normality indicate that the expectation for the disturbance terms $\epsilon_i$ in Equation~\ref{Method:E1} (e.g., sources associated with measurement error, stochastic effects, etc.) sum to 0 and we ignore these disturbance terms.

\subsection{Comparing estimates from BMA's optimal ensemble to SAMPL\-4 challenge methods} \label{Results:BMA_Methods}
The performance of all methods used in this work is shown in Table~\ref{Analysis:Table1:Methods} and Figure~\ref{Analysis:Figure1:Methods}.
Column 3 in Table~\ref{Analysis:Table1:Methods} lists the specific method for each estimation approach.
The performance results in column 4 of this table are consistent with results reported by Mobley et al. \cite{Mobley:2014}.
The performance of the optimal ensemble is shown in the last row.
The corresponding methods that are constituents in this optimal ensemble, ``alc-3'' and ``imp-2'', are highlighted in blue.
The final column in Table~\ref{Analysis:Table1:Methods} lists a direct comparison of each method's performance to the performance of the optimal ensemble: e.g., the ensemble reduces estimation errors by as much as 91\% in comparison to imp-6 and by 29\% in comparison to imp-2. 

\begin{figure}[h!]
	\centering
	\includegraphics[keepaspectratio,width=0.9\textwidth]{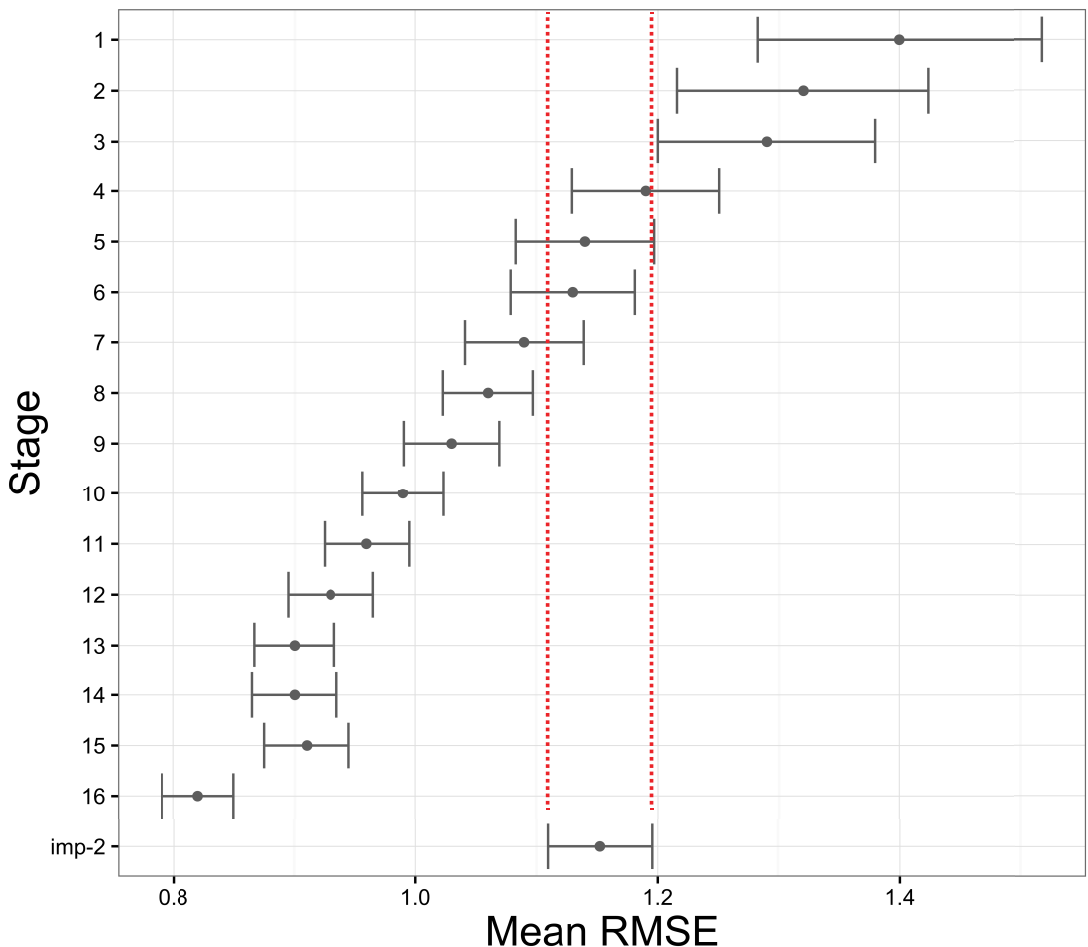}
	\caption{This plot illustrates the iterative pruning process discussed in Section~\ref{Method:StatEnsemble}. The y-axis depicts the mean root mean squared error (RMSE) of the different aggregated estimates based on ensembles formed during the different iterations of pruning. The x-axis indicates the stages of pruning. In general, \rev{the mean RMSE and the 95\% confidence intervals for each mean (based on standard error for the mean SEM) were reduced} with each iteration. The performance of the different ensembles are compared to the best performing method through the red line at $y=1.15$; all iterations past Stage 7 outperform the best method in the ensemble. Based on Wilcoxon generated p-values, the significance in the distributions of mean RMSE between the different ensembles are presented in Table~\ref{Analysis:Table2:BMA}. Based on mean RMSE, the optimal ensemble is created at Stage 16.}
	\label{Analysis:Figure2:BMA}
\end{figure}

The box plots in Figure~\ref{Analysis:Figure1:Methods} illustrate the performance variability across the different methods.
These plots also underscore the amount of uncertainty inherent in estimating solvation free energies: method-selection uncertainty plays a confounding factor in estimating solvation free energies.
Finally, the red line indicates the mean RMSE of the best-performing method: imp-2.
This line is also used in Figure~\ref{Analysis:Figure2:BMA} to contrast the iterative improvements obtained during the optimal ensemble's design process.
The ensemble's iterative design process based on Algorithms \ref{Method:Alg1} and \ref{Method:Alg2} is shown in Table~\ref{Analysis:Table2:BMA} and Figure~\ref{Analysis:Figure2:BMA}.
\begin{table}[t!]
	\centering
	\caption[Table 2]{This table lists the performance of the aggregated estimates obtained from the different ensembles created during the design process in Section~\ref{Method:StatEnsemble}.
	The second column lists the mean root mean squared error (RMSE) for each ensemble's aggregated estimate based on 100 iterations of a 2-fold cross-validation; this performance is also shown in Figure~\ref{Analysis:Figure2:BMA}.
	The third column lists the method that was selected to be pruned from the ensemble at the \emph{next} stage of the design process.
	The Wilcoxon generated p-values in column four are based on comparisons of mean RMSE distributions obtained from sequential ensembles and their aggregated estimates.
	$p$-values that are greater than $0.05$ are shown and indicate distributions that are equivalent (e.g., Stage 3 and 2) based on an $\alpha = 0.05$.
	Similarly, column five lists the Wilcoxon generated p-values reflecting comparisons between each stage and the best performing method, imp-2.
	As with column 4, bold $p$-values indicate that the performance between imp-2 and a given stage are equivalent (e.g., Stages 4-6).
	Based on mean RMSE and p-values from this table, the optimal ensemble is the one created in Stage 16; the final column in this table lists the performance improvement this ensemble provides in comparison to the ensembles generated in previous stages.}
	\scriptsize
	\rev{
	\begin{tabular}{l|c|c|c|c|c}
		\hline
		\hline
		Stage	& Ensemble RMSE		& Method 			& $p$-value		& $p$-value		& Improvement \\
				& (kcal mol$^{-1}$)	& Selected for		& (sequential 	& (stage vs.	& with BMA \\
				& Mean and Std Err Mean	& Pruning			& stages)		& imp-2)		& (Stage 16) \\
		\hline
		 1 & 1.40 $\pm$ 0.060 & imp-6 & NA & 0.000 & 41\%\\
		 2  & 1.32 $\pm$ 0.053 & alc-4&  \textbf{0.411} (vs. Stage 1) & 0.002 & 38\%\\
		 3 & 1.29 $\pm$ 0.046 & imp-8 & \textbf{1.000}  (vs. Stage 2) & 0.004 & 36\%\\
		 4 & 1.19 $\pm$ 0.031 & imp-5 & 0.012  (vs. Stage 3) & \textbf{0.356} & 31\%\\
		 5 & 1.14 $\pm$ 0.029 & imp-3 & 0.000  (vs. Stage 4) & \textbf{0.438} & 28\%\\
		 6 & 1.13 $\pm$ 0.026 & imp-7 & \textbf{0.883}  (vs. Stage 5)& \textbf{0.306} & 27\%\\
		 7 & 1.09 $\pm$ 0.025 & alc-5 & 0.015  (vs. Stage 6) &  0.027 & 25\%\\
		 8 & 1.06 $\pm$ 0.019 & alc-2 & \textbf{0.055}  (vs. Stage 7) & 0.003 & 23\%\\
		 9 & 1.03 $\pm$ 0.020 & imp-1 & \textbf{1.000}  (vs. Stage 8)& 0.001 & 20\%\\
		 10 & 0.99 $\pm$ 0.017 & imp-4 & 0.000  (vs. Stage 9)& 0.001 & 17\%\\
		 11 & 0.96 $\pm$ 0.018 & alc-1 & 0.000  (vs. Stage 10) & 0.000 & 15\%\\														
		 12 & 0.93 $\pm$ 0.018 & exp-4 &  \textbf{1.000}  (vs. Stage 11)& 0.000 & 12\%\\
		 13 & 0.90 $\pm$ 0.017 & hyb-2 & \textbf{1.000}  (vs. Stage 12) &0.000 &  09\%\\															
		 14 & 0.90 $\pm$ 0.018 & exp-1 & \textbf{1.000}   (vs. Stage 13)& 0.000 & 09\%\\															
		 15 & 0.91 $\pm$ 0.018 & exp-3 &   0.001 (vs. Stage 14)&  0.000  & 10\%\\																
		 16 & 0.82 $\pm$ 0.015 & NA  & 0.000   (vs. Stage 15)&0.000 &  0\%\\
		\hline
		\hline
	\end{tabular}}
	\label{Analysis:Table2:BMA}
\end{table}
The process starts with the 17 initial methods shown in Table~\ref{Analysis:Table1:Methods}.
Each subsequent row in Table~\ref{Analysis:Table2:BMA} represents an iteration through the pruning process.
The second column lists the mean \rev{RMSE} of each stage's ensemble based on its aggregated estimate.
At the end of each stage, a method is selected to be pruned from the existing ensemble before proceeding to the next iteration; the specific method that was selected is shown in column 3.
For example, at Stage 1 there were 17 methods in the ensemble and imp-6 was selected to be removed.
During the next step, Stage 2, imp-6 \rev{was} removed from the ensemble \rev{so that only 16 methods were used to create the aggregated estimate}.
At the end of this stage, alc-4 was selected to be pruned for Stage 3.
In the final stage, the only remaining methods in the ensemble were imp-2 and alc-3.
This iterative design process is also graphically illustrated in Figure~\ref{Analysis:Figure2:BMA} showing how selective pruning increases the performance of each successive ensemble.
The red line in this figure indicates the performance of the best performing method, imp-2.
The benefits of the aggregated estimates become apparent after Stage 6 where the ensembles outperform imp-2.

The statistical significance of the iterative design process is listed in columns 4 and 5 in Table~\ref{Analysis:Table2:BMA}; bold $p$-values in these columns indicate the performance between two distributions are equivalent.
Column 4 lists the Wilcoxon generated $p$-values based on comparisons of mean RMSE distributions obtained from sequential ensembles.
For example, in the second row the $p$-value for the comparison of Stage 2 vs.~Stage 1 indicates that these distributions are equivalent.
Contrariwise, in row 4, the $p$-value for the comparison between Stage 4 and Stage 3 indicates that the distributions are not equivalent; the mean RMSE in column 2 indicates that the ensemble of Stage 4 outperforms the ensemble built in Stage 3.

As a second analysis of significance, column 5 lists the Wilcoxon-generated $p$-values that represent comparisons between each ensemble and the best-performing method, imp-2.
From these values, Stages 4-6 are seen to be equivalent to imp-2.
However, the mean RMSE \rev{distributions} of all subsequent stages are not equivalent: based on mean RMSE listed in column 2, we conclude that these successive ensembles (increasingly) outperform imp-2.
The p-values in these columns and the mean RMSE results demonstrate that the optimal ensemble is created in Stage 16.
The final column in Table~\ref{Analysis:Table2:BMA} lists the performance improvement that the Stage 16 ensemble provides in comparison to \rev{all the other ensembles generated in the design process}.

\begin{figure}[h!]
	\centering
	\includegraphics[keepaspectratio,width=0.9\textwidth]{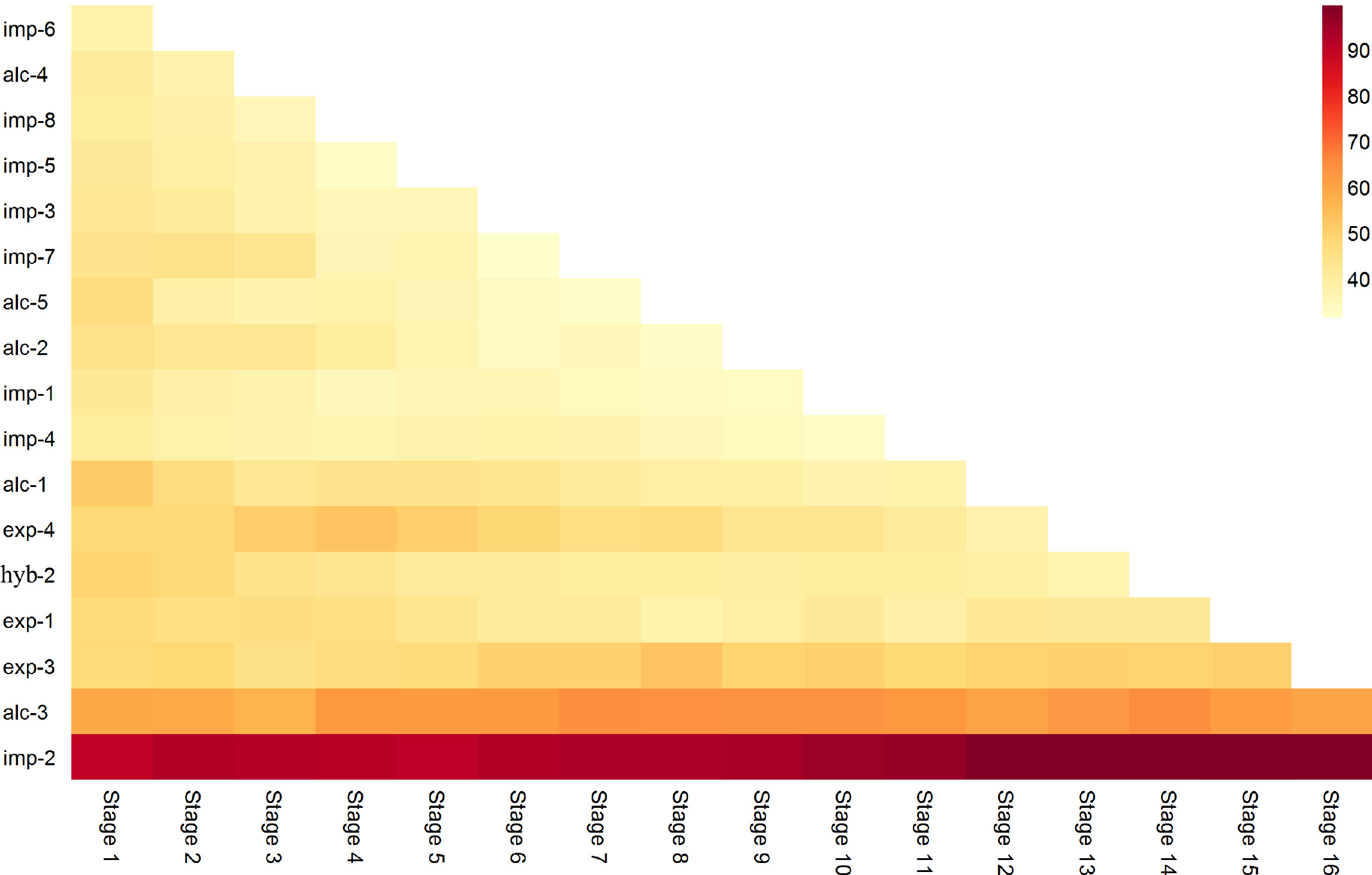}
	\caption{This image provides a graphical depiction of the ensemble design process performed on the initial ensemble of 17 methods.
	The color scale represents the probability, 0 - 100\%, that a given method's coefficient term, $\beta_j$ will not be zero.
	The stages in the design process, starting at the left and moving to the right, are shown on the x-axis.
	The methods in each ensemble are shown on the y-axis; note that the methods are listed (from top to bottom) in the order that they are pruned in the design process.
	Thus at Stage 1, all methods are used in the ensemble and their $\mathrm{Pr}(\beta_j^{\text{BMA}}\neq 0) $ range from ~40\% (e.g., imp-6) to ~90\% (imp-2).
	By Stage 3, imp-6 and alc-4 have been pruned from the ensemble and the probability values have adjusted accordingly as shown in the colored column above Stage 3.
	In general the trend across the different stages of pruning illustrates that methods become increasingly lighter, i.e., the ensemble design process becomes increasingly confident that these methods (e.g., imp-7 and exp-4) are not needed in the ensemble.
	\rev{Contrariwise, the $\mathrm{Pr}(\beta_j^{\text{BMA}}\neq 0) $ for a few methods (e.g., alc-3, and imp-2) remain above 70\% and even increase throughout the design process, indicating a high degree of confidence in the statistical significance of these methods.}}
	\label{Analysis:Figure3:P-not}
\end{figure}

Figure~\ref{Analysis:Figure3:P-not} depicts a heat-map that shows the statistical information that drove the pruning process for each stage.
In this image, the y-axis represents the different methods; the x-axis (starting from left) indicates the successive stages in the design process.
The color scale represents the mean probability, 0 - 100\%, that a given method's coefficient term, $\beta_j$ will not be zero.
At Stage 1, all methods were used in the ensemble and their $\mathrm{Pr}(\beta_j^{\text{BMA}}\neq 0)$ ranged from~40\% (imp-6) to~80\% (imp-2).
As imp-6 had the lowest mean probability of not being zero, imp-6 was pruned after Stage 1; the color map colors this method white in Stage 2 to indicate it has been eliminated.
The methods listed on the legend of the y-axis are  ordered by the sequence that they were eliminated in the design process: imp-6 first, alc-4 second, imp-8 third, etc.

\subsection{Performance \rev{analysis based on compounds}}
\label{Results:BMA_Molecules}
\begin{figure}
	\centering
	\includegraphics[keepaspectratio,width=0.75\textwidth]{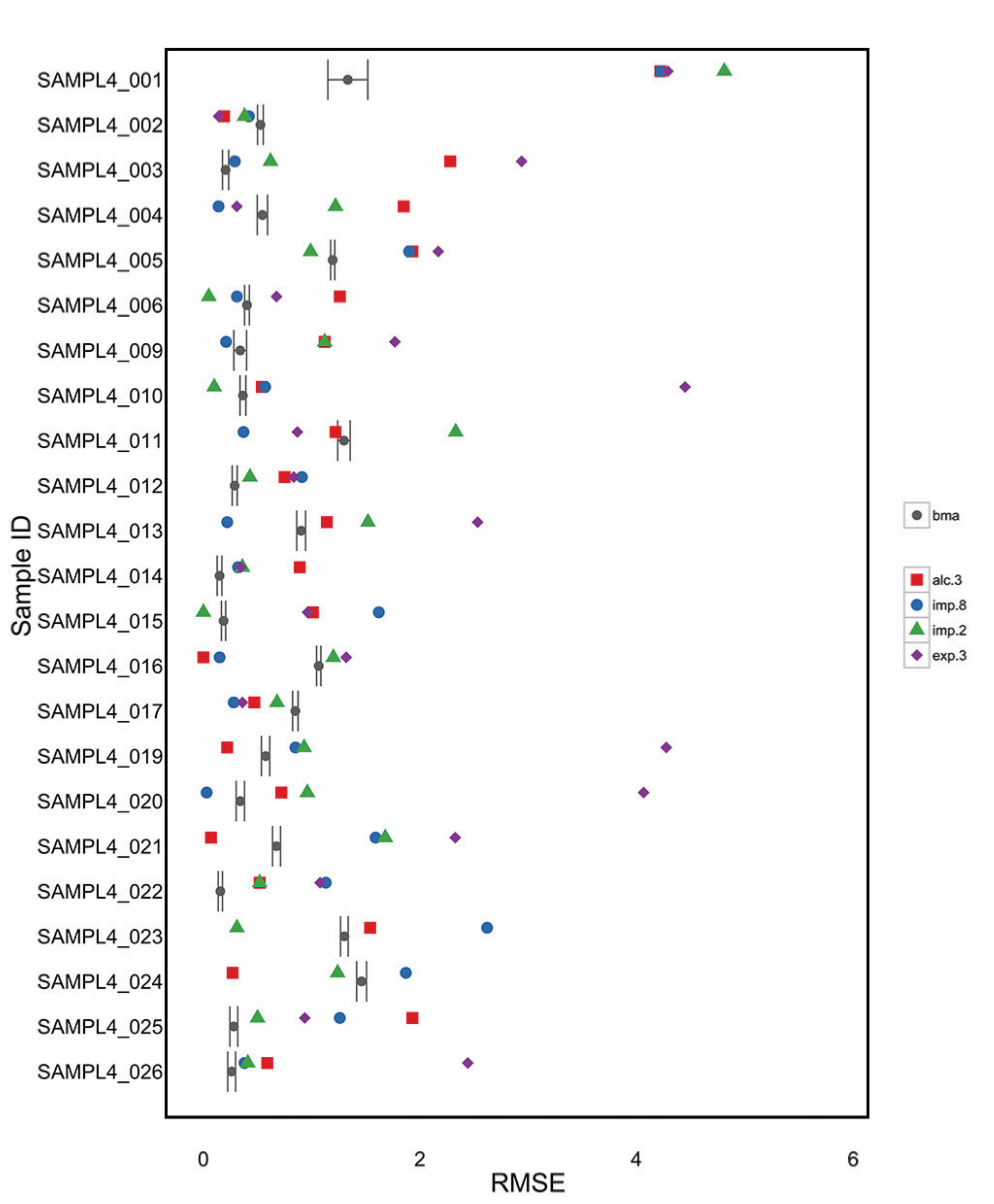}
	\caption{This figure is one of two figures (Figures~\ref{Analysis:Figure4:Compounds1} and~\ref{Analysis:Figure5:Compounds2}) that depict the RMSE (kcal mol$^{-1}$) performance of several methods based on the individual compounds taken from the SAMPL4 challenge: the first, second, and third \rev{best-}performing methods (i.e., imp-2, imp-8, and alc-3) as well as exp-3.
	Note that imp-2 and alc-3 are the methods used in the optimal BMA ensemble (Stage 15) and exp-3 is the final method eliminated from the ensemble (Table~\ref{Analysis:Table2:BMA}) in Stage 16.
	BMA's performance based on the optimal ensemble is shown based on its distribution of the mean \rev{RMSE} for estimates made in our 2-fold cross-validation analysis.
	Of note is the performance of the different methods for SAMPL4\_022 (mefenamic acid), SAMPL4\_023 (diphenhydramine), SAMPL4\_027 (1,3-bis-(nitroxy)propane), SAMPL4\_009 (2,6-dichlorosyringaldehyde), and SAMPL4\_001 (mannitol).
	These are the most challenging compounds for methods to estimate based on the analysis of Mobley et al.\@ of the SAMPL4 data~\cite{Mobley:2014}.
	The benefits of the ensemble is clearly demonstrated here as the BMA ensemble outperforms all methods in estimating SAMPL4\_022, SAMPL4\_009, and SAMPL4\_001.
	For SAMPL4\_023 and SAMPL4\_027 the ensemble provides the second best performance, and in this context provides more consistent performance than the other methods: e.g., alc-3 is better at estimating SAMPL4\_027, but is third at estimating SAMPL4\_23.}
\label{Analysis:Figure4:Compounds1}
\end{figure}

\begin{figure}
	\centering
	\includegraphics[keepaspectratio,width=0.75\textwidth]{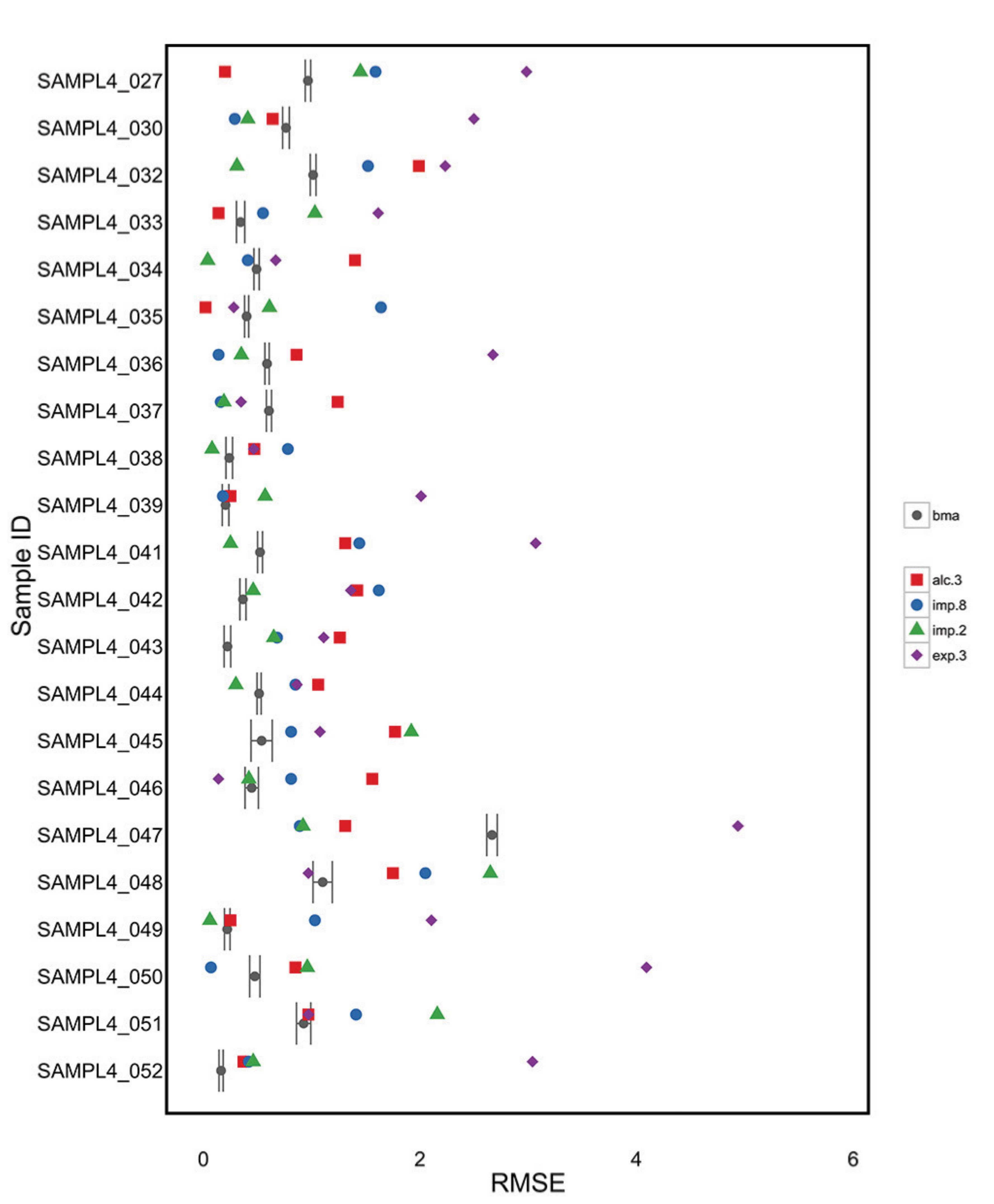}
	\caption{This figure is one of two figures (Figures~\ref{Analysis:Figure4:Compounds1} and~\ref{Analysis:Figure5:Compounds2}) that depict the RMSE (kcal mol$^{-1}$) performance of several methods based on the individual compounds taken from the SAMPL4 challenge: the first, second, and third best-performing methods (i.e., imp-2, imp-8, and alc-3) as well as exp-3.
	See Figure \ref{Analysis:Figure4:Compounds1} for more information about both plots.}
	\label{Analysis:Figure5:Compounds2}
\end{figure}

Figures~\ref{Analysis:Figure4:Compounds1} and~\ref{Analysis:Figure5:Compounds2} depict \rev{the} performance of different methods according to specific SAMPL4 challenge small molecule compounds.
In addition to the optimal ensemble, we also show the performance of the first, second, and third best-performing methods from the SAMPL4 challenge: imp-2, imp-8, and alc-3.
Methods imp-2 and alc-3 are the methods used in the optimal BMA ensemble and exp-3 is the final method eliminated from the ensemble (Stage 15 in Table~\ref{Analysis:Table2:BMA}). 

\rev{Mobley et al.~\cite{Mobley:2014} note that prediction error generally increases as molecular complexity and size increases.} 
Specifically, polyfunctional molecules with several interacting groups were especially difficult to estimate in the SAMPL4 challenge: SAMPL4\_022 (mefenamic acid), SAMPL4\_023 (diphenhydramine), SAMPL4\_027 (1,3-bis-(nitroxy)propane), SAMPL4\_009 (2,6-dichlorosyringaldehyde), and SAMPL4\_001 (mannitol). 
Figures~\ref{Analysis:Figure4:Compounds1} and~\ref{Analysis:Figure5:Compounds2} show that the optimal ensemble outperforms all methods in estimating SAMPL4\_022 and SAMPL4\_001, and provides the second best estimates for SAMPL4\_009, SAMPL4\_023 and SAMPL4\_027.
Aside from the optimal ensemble, there is no clear best method for estimating these compounds.
For example, while imp-8 is best at estimating SAMPL4\_009, it is does not do well at estimating either SAMPL4\_023 or SAMPL4\_027.
Similarly while imp-2 performs well at estimating SAMPL4\_023, it does not do as well at estimating SAMPL4\_009 or SAMPL4\_027.
The general trend in performance \rev{indicates} that the optimal ensemble provides more consistent and accurate estimates than any specific method. 

\subsection{\rev{Interpreting aggregate estimate performance}}
\rev{Aggregate estimate performance is degraded by both variable (i.e., inconsistent) predictive performance of its method constituents and the uncertainties associated with statistical model construction.
These sources of error manifest in different ways and we can characterize trends in estimation accuracy and consistency to support our analysis of these sources.}

\rev{If methods are sensitive to functional discrepancies between training and testing molecules, then aggregate estimates that rely on these methods may perform poorly on chemical  compounds that are outside the training set. 
While some of these sensitivities may be obvious to identify (e.g., methods that are only accurate for certain chemical classes), most are much more subtle and difficult to detect. 
Cross-validation tests, like those performed in Section~\ref{Method:StatEnsemble:Design}, help assess the more obvious class or functional group-based sensitivities.
More specifically, under cross-validation tests the overall performance of the aggregate estimate will be significantly poorer in comparison to other methods. 
This trait will be most notable when comparing the aggregate against its method constituents on molecules where constituents are very accurate.
In general, as method sensitivity increases, estimate accuracy and eventually consistency are negatively impacted.}

\rev{The overall performance of the aggregate estimate is better than its constituents (Table~\ref{Analysis:Figure1:Methods}, Figures~\ref{Analysis:Figure4:Compounds1} and~\ref{Analysis:Figure5:Compounds2},) and significantly better than other methods.
Based on these observations, large method-based sensitivities do not appear to be an issue for the SAMPL4 challenge data and so we next proceed to assess
the sensitivities associated with statistical model construction.}

\rev{Statistical sensitivities in a model manifest in estimate consistency in a cross-validation study.
In this work, the 95\% confidence intervals of the expected mean provide a quantitative metric for assessing these types of sensitivities. 
Under the assumption of minimal method sensitivity, a perfect statistical model will provide very precise and accurate estimates, regardless of training and testing data.
The further away a statistical model is from this hypothetical ``perfect'' model, the more uncertainty will impact how weights are assigned in its coefficient vector, $\beta_j$.
As a result, changes in training or testing data will results in an increasingly wider distribution of estimates (i.e., wider confidence intervals) and will impact the model's accuracy.
In extreme cases, these distributions may even become bimodal.}

\rev{The overall performance of the aggregate estimate is very consistent across chemical compounds (Figures~\ref{Analysis:Figure4:Compounds1} and~\ref{Analysis:Figure5:Compounds2},) with the exception of SAMPL4\_001 and SAMPL4\_045.
For SAMPL4\_001 and SAMPL4\_045, the 95\% confidence intervals for the expected mean are noticeably wider then other compounds.
As these estimates are still very accurate in comparison to the ensemble's constituents, it is reasonable to assume that the underlying statistical model of the optimal is largely valid across the space of chemical compounds defined by the SAMPL4 challenge.
Thus, while there appears to be indications of statistical sensitivity, it may be more effective to address the performance of methods for these two compounds first before refining the statistical model.}

\rev{Given the assumption of minimal method and statistical sensitivities, there are several chemical compounds that deserve specific attention: SAMPL4\_002, SAMPL4\_017, SAMPL4\_024, SAMPL4\_030, and SAMPL4\_047.
In all of these cases, aggregate's estimate is outperformed by its constituents.
For three of the compounds,  SAMPL4\_002, SAMPL4\_017, and SAMPL4\_030, the degree of error is very small in comparison to the ensemble's constituents. 
It is likely that statistical uncertainties are playing a role in degrading performance for these three such that the underlying model has lost some accuracy, but maintained consistency.
Addressing these concerns is usually best done by increasing the size of training and testing pools to provide the statistical design process more exemplar data.}

\rev{In comparison to the previous compounds, aggregate estimate performance for SAMPL4\_024 and SAMPL4\_047 is notably poor.
Further, the usual indicators of method and statistical sensitivities do not provide evidence to explain this performance loss.
For example confidence bounds for these estimates are narrow, indicating that estimates are consistent regardless of the cross-validation training and testing sets.
Additionally, total mean estimation errors are largely restricted to these two specific chemicals, indicating that overall the model is valid and accurate for the SAMPL4 challenge data.}
 
\rev{One explanation for these trends may be that the current optimal ensemble experiences a more subtle statistical or method-based sensitivity for these specific chemical compounds. 
For example, Mobley et al.~\cite{Mobley:2014} indicate that these two specific chemical compounds are the largest MW compounds evaluated by the SAMPL4 challenge: SAMPL4\_024 is amitriptyline (MW 277.41) and SAMPL4\_047 is 1-(2-hydroxyethylamino)-9,10-anthraquinone (MW 267.28).
Additionally, SAMPL4\_047 is listed as the second largest hydration free energy in the challenge.
It is thus feasible that the ensemble design process has identified a statistical model that is valid for lighter weight, lower hydration free energy compounds but is not sufficiently accurate for larger compounds with higher free energy values.
To test this hypothesis, the ensemble design process would need to expand to include more chemical exemplars that are similar to SAMPL4\_024 and SAMPL4\_047.}

\rev{This statistical framework has designed a single model for combining two methods in order to estimate hydration free energies for any compound in the SAMPL4 challenge.
These method constituents are based on two different approaches for predicting hydration free energy: one is based on a multi-conformational implicit method and requires tens of minutes to make a prediction, where as the other is based on molecular dynamics and requires many hours to make the same prediction. 
One benefit of this approach therefore is not only in improving estimation accuracy, but in providing this increase in accuracy by combining computationally expensive methods with cheaper methods.
We envision that through this type of framework, specific ensembles of methods will be able to be developed for different classes of chemical compounds.
In this way, a stratification of the chemical landscape (e.g., based on MW, functionality and polarization, etc.) is possible where specific classes of  compounds are mapped to distinct ensembles, and these ensembles are able to provide the best estimates for hydration free energies at the lowest possible computational cost.}

\subsection{\rev{Performance analysis of alternate ensemble techniques}}
\label{Results:BMA_Variants}
\rev{There are alternative approaches to BMA that can combine an ensemble of methods into an aggregate estimate.
The notion of method and statistical sensitivities are also points of concern for these alternate approaches.}
In our cross-validation study, we evaluated four common approaches for aggregating an ensemble and evaluated their estimated benefits in comparison to BMA.
These methods are listed in Table~\ref{Analysis:Table4:EnsembleCompare} and include: Random Forest~\cite{Breiman:2001}, Ridge Regression~\cite{Hoerl:2000}, Lasso~\cite{Tibshirani:1994}, and stepwise regression via forward selection.

\rev{The difference between the BMA approach and these other approaches is in their design process.
More specifically, each approach has a strategy for performing model specification (e.g., Section~\ref{MethodMain}).
As we have addressed the notion of method sensitivities in the last section, the objective for this section is to compare these different approaches for model construction to highlight the effects of model sensitivities.
As each method will select its own methods to form its own ensemble, the difference in performance for these approaches will highlight the benefits of addressing model sensitivities.
In other words, relying on a modeling approach that does not fully address these sensitivities can mislead analysis of ensemble performance and thwart subsequent ensemble design.}

\rev{The random forest approach is an ensemble learning method similar to BMA.
	It operates by constructing multiple decision trees at training time and outputs an estimate based on the mean prediction of the individual trees. 
Where as the random forest algorithm relies on the non-informative weighting, BMA's approach relies on a Bayesian weighting for each estimate.
The random forest approach is well established in computational biology and bioinformatic applications.}

\rev{The ridge regression approach is a technique for analyzing regression data that may suffer from multicollinearity. 
When multicollinearity occurs, it can create inaccurate estimates of the regression coefficients, $\beta_j$, resulting in an increase of standard error.
Ridge regression addresses the concerns for high variance by adding a degree of bias to the regression estimates that will reduce overall estimate errors.} 

\rev{Lasso and forward-selection are both strategies for selecting methods to use in a linear regression model (Equation~\ref{Method:E1}). 
In forward-selection, the approach starts with the assumption that all coefficients, $\beta_j$ are equal to zero.
Next, it selects the method, $x_ij$ most correlated with $y_i$, and adds it into the model. 
It then assesses the performance of the model and continues to add the next highest correlated methods to the model until performance no longer increases.
Lasso follows the same general approach, but doesn't add a method fully into the model.
Instead, Lasso penalizes the absolute size of the regression coefficients until the coefficient, $\beta_j$ for the method being added is no longer the one most correlated method with overall performance.}

\rev{Table~\ref{Analysis:Table5:EnsembleVariables} lists the methods that were selected by each modeling approach. From largest to smallest ensemble the modeling ensembles are: Random Forest 14, Ridge Regression 7, Lasso 4, Forward Selection 4, and BMA 2.}

\begin{table}[t!]
	\centering
	\caption[Methods used by each modeling approach]{\rev{This table lists the different methods used by each modeling approach in Section~\ref{Results:BMA_Variants}.  The performance for each of these methods is listed in Table~\ref{Analysis:Table4:EnsembleCompare} and Figure~\ref{Analysis:Figure6:Models}. Starting from the top and working down, the total number of methods used in each ensemble are: Random Forest 14, Ridge Regression 7, Lasso 4, Forward Selection 4, and BMA 2.}}
	\footnotesize
	\rev{
	\begin{tabular}{lccccccccc}
		\hline
		\hline
		Model  & imp-1 & alc-5 & imp-4 & exp-3 & imp-5 & alc-2 & exp-1 & exp-4 & alc-4\\
		\hline
		Random Forest & X & X & - & X & - & X & X & X & -\\
		Ridge & - & X & - & - & - & X & X & - & -\\
		Lasso & - & - & - & - & - & - & - & - & -\\
		Forward Selection & - & - & - & - & - & - & - & X & -\\
		BMA (Stage 16) & - & - & - & - & - & - & - & - & -\\		
		\hline
		\\
		\hline
		Model & imp-7 & hyb-2 & alc-1 & imp-3 & imp-1 & alc-3 & imp-8 & imp-2 &\\
		\hline
		Random Forest & X & X & X & X & X & X & X & X &\\
		Ridge & - & - & X & X & - & X & - & X &\\
		Lasso & - & - & X & X & - & X & - & X &\\
		Forward Selection & - & - & - & X & - & X & - & X &\\
		BMA (Stage 16) & - & - & - & - & - & X & - & X &\\		
		\hline
		\hline
	\end{tabular}}
	\label{Analysis:Table5:EnsembleVariables}
\end{table}

\begin{figure}
	\centering
	\includegraphics[keepaspectratio,width=0.9\textwidth]{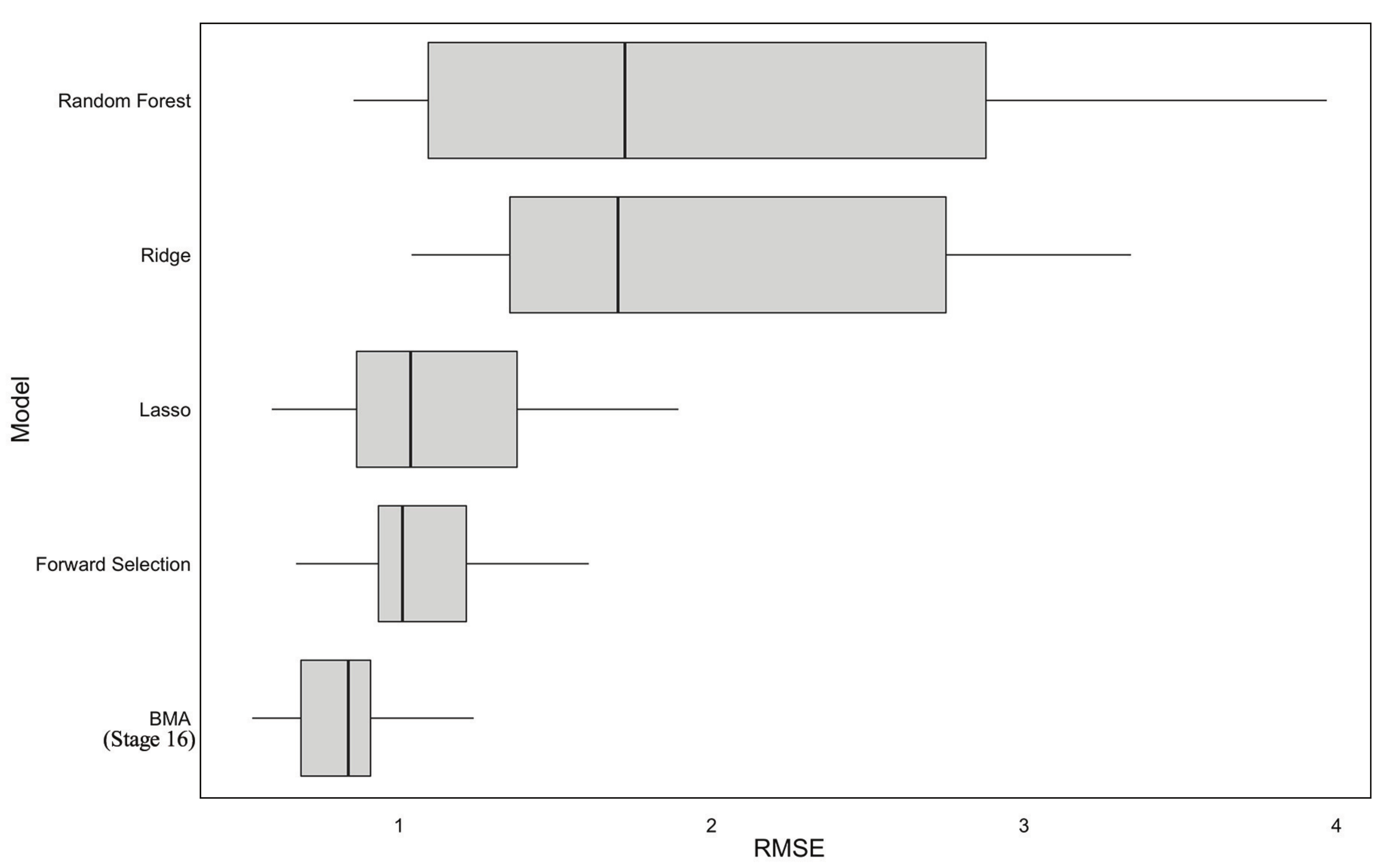}
	\caption{This figure displays the performance of different ensemble approaches in comparison to the optimal ensemble designed in this work, BMA (Stage 16).
	The mean \rev{RMSE}, min, max, first, and third quartiles (kcal mol$^{-1}$) of these ensemble approaches are shown based on the 100 iterations of the 2-fold cross-validation experiment discussed in Section~\ref{Method:StatEnsemble}.
	Based on an $\alpha = 0.05$, the Wilcoxon based p-values (Table~\ref{Analysis:Table4:EnsembleCompare}) indicate that BMA's improved performance is statistically significant to the other approaches that can combine an ensemble of methods to make an aggregated estimate.}
	\label{Analysis:Figure6:Models}
\end{figure}

\begin{table}[t!]
	\centering
	\caption[Model comparison with mean RMSE]{This table lists the performance of different ensemble approaches in comparison to the optimal ensemble designed in this work through BMA.
	The performance of these ensemble approaches, given as the mean \rev{RMSE} with standard deviation, are based on the 100 iterations of the 2-fold cross-validation experiment discussed in Section~\ref{Method:StatEnsemble}.
	Based on an $\alpha = 0.05$, the Wilcoxon based $p$-values indicate that BMA's improved performance is statistically significant to the other ensemble approaches for combining methods to make an aggregated estimate.
	The last column indicates the improvement in estimation that the optimal ensemble provides to these alternate techniques: estimation accuracy is improved from 25\% to 61\%. }
	\footnotesize
	\begin{tabular}{lccc}
		\hline
		\hline
		Ensemble Model  & Ensemble Mean RMSE  & p-value & Improvement\\
		 & and Std \rev{Err Mean}& & by BMA (Stage 16)\\
		\hline
		Random Forest & 2.08 $\pm$1.00 & 0.00 & 61\%\\
		Ridge & 2.06 $\pm$0.75 & 0.00 & 60\%\\
		Lasso & 1.12 $\pm$0.34 & 0.00 & 27\%\\
		Forward Selection & 1.09 $\pm$0.26 & 0.00 &  25\%\\		
		BMA (Stage 16) & 0.82 $\pm$0.17 & NA & 0\%\\
		\hline
		\hline
	\end{tabular}
	\label{Analysis:Table4:EnsembleCompare}
\end{table}
These techniques were chosen as they have all been used successfully for a variety of inference tasks and are readily available for use~\cite{R:2008,sklearn_api:2013}.
\rev{As all of these approaches construct an estimate for $\beta$, training and estimating with these approaches was performed identically to BMA, using the cross-validation detailed in Section~\ref{Method:StatEnsemble}.
More specifically, the performance reflects the result of 100-iterations of a 2-fold cross-validation.
We also followed the same procedure for comparing BMA's estimating capability to these alternate ensemble-based techniques.}
Figure~\ref{Analysis:Figure6:Models} and Table~\ref{Analysis:Table4:EnsembleCompare} provide an overview of the cross-validation errors for the various ensemble-based approaches and the BMA-based optimal ensemble.
The statistical significance of BMA's performance in Figure~\ref{Analysis:Figure6:Models} is based on $p$-values shown in Table~\ref{Analysis:Table4:EnsembleCompare}. Based on an $\alpha = 0.05$, Table \ref{Analysis:Table4:EnsembleCompare} indicates that we reject the null hypothesis for all paired comparison tests.
BMA's mean RMSE distribution is therefore not equivalent to the mean RMSE distribution of any other ensemble-based technique. 
As the distributions are not equal, we compared mean RMSE distributions of BMA to the other ensemble-based approaches in Figure~\ref{Analysis:Figure6:Models} and Table~\ref{Analysis:Table4:EnsembleCompare}.
From these mean RMSEs, it is clear that the BMA-based approach outperforms all other ensemble-based \rev{estimation} approaches: BMA-based estimates reduced error by approximately 60\% in comparison to Random Forest and Ridge regression methods. In comparison to Lasso, BMA reduces estimation error by approximately 27\%.
Finally, in comparison to stepwise regression via forward selection, BMA reduces error by approximately 25\%.

\section{Conclusions}
This study demonstrates a proof-of-principle application of how to statistically design and aggregate an ensemble of methods for estimating solvation free energies in small molecules.
While the performance of BMA is expected to generalize to a much broader set of small molecule estimation problems, the specific BMA model trained in this study is likely to be dependent on the small molecules used in the SAMPL4 challenge.

\rev{In future work, we will investigate the application of this method to broader regions of chemical space.
Specifically, we envision that this type of framework can help stratify the small molecule chemical landscape based on chemical classes defined by molecular weight, functionality and polarization, available surface area, etc.
This stratification will map specific classes of compounds to distinct ensembles that provide the best estimates for hydration free energies at the lowest possible computational cost.
Towards this task, we are looking at penalizing computationally expensive methods that provide minimal accuracy benefits to optimize both the accuracy of the estimates and the efficiency of the calculation.}

\subsection*{Acknowledgments}
\rev{We thank the anonymous reviewers for their helpful questions and suggestions, and the SAMPL participants and organizers for making the data available.} 
This research was funded by NIH grant R01 GM069702.

\bibliography{refs}
\bibliographystyle{unsrt}


\end{document}